\documentclass[doublecol]{epl2} 
\usepackage{graphicx,amsfonts,amssymb,amsmath,hyperref}
\usepackage{textcomp}
\usepackage{esint}

\title{Quantum XY criticality in a two-dimensional Bose gas near the Mott transition}
\shorttitle{XY quantum criticality in a two-dimensional Bose gas} 

\author{A. Ran\c{c}on\inst{1,2} \and N. Dupuis\inst{2}}
\shortauthor{A. Ran\c{c}on \etal}

\institute{                    
  \inst{1} James Franck Institute and Department of Physics,
University of Chicago, Chicago, Illinois 60637, USA  \\
  \inst{2} Laboratoire de Physique Th\'eorique de la Mati\`ere Condens\'ee, 
CNRS UMR 7600, Universit\'e Pierre et Marie Curie, 4 Place Jussieu, 
75252 Paris Cedex 05, France
}
\pacs{67.85.-d}{Ultracold gases, trapped gases}
\pacs{05.30.Rt}{Quantum phase transitions}
\pacs{05.30.Jp}{Boson systems}

\abstract{
We derive the equation of state of a two-dimensional Bose gas in an optical lattice in the framework of the Bose-Hubbard model. We focus on the vicinity of the multicritical points where the quantum phase transition between the Mott insulator and the superfluid phase occurs at fixed density and belongs to the three-dimensional XY model universality class. Using a nonperturbative renormalization-group approach, we compute the pressure $P(\mu,T)$ as a function of chemical potential and temperature. Our results compare favorably with a calculation based on the quantum O(2) model -- we find the same universal scaling function -- and allow us to determine the region of the phase diagram in the vicinity of a  quantum multicritical point where the equation of state is universal. We also discuss the possible experimental observation of quantum XY criticality in a ultracold gas in an optical lattice.}

\begin{document}

\maketitle

\def\rhoeq{\hat\rho_{\rm eq}}

\newcommand{\marge}[1]{\marginpar{\scriptsize #1}}
\newcommand{\remarque}[1]{\marginpar{\scriptsize Remarque}{\it [#1]}}
\newcommand{\new}[1]{{\bf #1}}

\def\beq{\begin{equation}}
\def\eeq{\end{equation}}
\def\bleq{\begin{eqnarray}}
\def\eleq{\end{eqnarray}} 
\def\bfig{\begin{figure}}
\def\efig{\end{figure}}
\def\bline{\begin{multline}}
\def\eline{\end{multline}}
\def\bremark{\begin{quotation} \noindent \small }
\def\eremark{\end{quotation}}
\def\llbrace{\left\lbrace}
\def\rrbrace{\right\rbrace}
\def\lbraket{\left[}
\def\rbraket{\right]}

\newcommand{\Tr}{{\rm Tr}} 
\newcommand{\tr}{{\rm tr}} 
\newcommand{\sgn}{{\rm sgn}} 
\newcommand{\mean}[1]{\langle #1 \rangle}
\newcommand{\commu}[2]{[#1,#2]} 
\newcommand{\bra}[1]{\langle#1|}
\newcommand{\ket}[1]{|#1\rangle}
\newcommand{\braket}[2]{\langle #1|#2\rangle}
\newcommand{\dbraket}[3]{\langle #1|#2|#3\rangle}
\newcommand{\vac}{|{\rm vac}\rangle} 
\def\bravac{\langle{\rm vac}|}
\newcommand{\const}{{\rm const}} 
\newcommand{\atanh}{\,{\rm atanh}}

\newcommand{\ie}{i.e. }
\newcommand{\iet}{i.e.}
\newcommand{\eg}{e.g. }
\newcommand{\cc}{{\rm c.c.}} 
\newcommand{\hc}{{\rm h.c.}} 
\def\etal{{\it et al. }}

\newcommand{\jhatbf}{\hat {\textbf \j}} 
\newcommand{\Jhatbf}{\hat {\textbf \J}} 
\newcommand{\jhat}{\hat {\jmath}} 
\newcommand{\Jhat}{\hat {J}} 
\newcommand{\jbf}{\textbf j}
\newcommand{\Jbf}{\textbf J}

\def\chibf{\boldsymbol{\chi}}
\def\down{\downarrow}
\def\eps{\epsilon}
\def\gam{\gamma} 
\def\phibf{\boldsymbol{\phi}}
\def\varphibf{\boldsymbol{\varphi}}
\def\varphibfs{\boldsymbol{\varphi}_<}
\def\varphibfl{\boldsymbol{\varphi}_>}
\def\varphis{\varphi_{<}}
\def\varphil{\varphi_{>}}
\def\psibf{\boldsymbol{\psi}}
\def\Ome{\Omega}
\def\omeD{{\omega_D}} 
\def\bfOme{\boldsymbol{\Omega}} 
\def\Omebf{\boldsymbol{\Omega}} 
\def\lamb{\lambda}
\def\Lamb{\Lambda}
\def\sig{\sigma}
\def\sigp{{\sigma'}} 
\def\bfsig{\boldsymbol{\sigma}} 
\def\sigbf{\boldsymbol{\sigma}} 
\def\The{\Theta} 
\def\up{\uparrow}

\def\epsk{\epsilon_{\bf k}} 
\def\xik{\xi_{\bf k}} 
\def\xikq{\xi_{{\bf k}+{\bf q}}} 
\def\Ek{E_{\bf k}}
\def\Heff{\hat H_{\rm eff}}
\def\Hem{\hat H_{\rm em}}
\def\Hint{\hat H_{\rm int}}
\def\Hloc{\hat H_{\rm loc}}
\def\HMF{\hat H_{\rm MF}}
\def\Sem{S_{\rm em}}
\def\SMF{S_{\rm MF}} 
\def\SRPA{S_{\rm RPA}} 
\def\Sint{S_{\rm int}} 
\def\Sloc{S_{\rm loc}} 
\def\Zloc{Z_{\rm loc}} 
\def\ZMF{Z_{\rm MF}} 
\def\ZRPA{Z_{\rm RPA}} 
\def\RPA{{\rm RPA}}
\def\loc{{\rm loc}} 
\def\pp{{\rm pp}}
\def\ph{{\rm ph}} 
\def\ch{{\rm ch}}
\def\sp{{\rm sp}} 
\def\qtf{q_{\rm TF}}
\def\epstf{\eps^{}_{\rm TF}} 
\def\epsrpa{\eps^{}_{\rm RPA}} 
\def\chinnzpp{\chi_{nn}^{0}{}\!\!\!''}

\def\half{\frac{1}{2}}
\def\dhalf{\dfrac{1}{2}}
\def\third{\frac{1}{3}} 
\def\quarter{\frac{1}{4}}

\def\qr{{\bf q}\cdot{\bf r}}
\def\wt{\omega t} 

\def\a{{\bf a}}
\def\b{{\bf b}}
\def\e{{\bf e}}
\def\f{{\bf f}}
\def\g{{\bf g}}
\def\h{{\bf h}}
\def\k{{\bf k}}
\def\l{{\bf l}}
\def\m{{\bf m}}
\def\n{{\bf n}} 
\def\p{{\bf p}} 
\def\q{{\bf q}}
\def\r{{\bf r}}
\def\t{{\bf t}}
\def\u{{\bf u}}
\def\v{{\bf v}}
\def\x{{\bf x}}
\def\y{{\bf y}} 
\def\z{{\bf z}} 
\def\A{{\bf A}}
\def\B{{\bf B}}
\def\D{{\bf D}} 
\def\E{{\bf E}} 
\def\F{{\bf F}} 
\def\H{{\bf H}}  
\def\J{{\bf J}}
\def\K{{\bf K}} 

\def\G{{\bf G}}
\def\L{{\bf L}}
\def\M{{\bf M}}  
\def\O{{\bf O}} 
\def\P{{\bf P}} 
\def\Q{{\bf Q}} 
\def\R{{\bf R}}
\def\S{{\bf S}}
\def\epsbf{\boldsymbol{\epsilon}}
\def\mubf{\boldsymbol{\mu}}
\def\nablabf{\boldsymbol{\nabla}}
\def\rhobf{\boldsymbol{\rho}}
\def\sigmabf{\boldsymbol{\sigma}} 
\def\Pibf{\boldsymbol{\Pi}}
\def\pibf{\boldsymbol{\pi}}

\def\para{\parallel}
\def\kpara{{k_\parallel}}
\def\kperp{{k_\perp}} 
\def\kperpp{{k_\perp'}} 
\def\qperp{{q_\perp}} 
\def\tperp{{t_\perp}} 

\def\w{\omega}
\def\wn{\omega_n}
\def\wnu{\omega_\nu}
\def\wp{\omega_p} 
\def\dmu{{\partial_\mu}}
\def\dl{{\partial_l}}  
\def\dt{\partial_t} 
\def\tdt{\tilde\partial_t}
\def\dk{\partial_k}
\def\tdk{\tilde\partial_k}
\def\dx{\partial_x}
\def\dy{\partial_y} 
\def\dtau{{\partial_\tau}}  
\def\det{{\rm det}} 
\def\Pf{{\rm Pf}}

\def\dsum{\displaystyle \sum}
\def\dint{\displaystyle \int} 
\def\intt{\int_{-\infty}^\infty dt} 
\def\inttp{\int_{-\infty}^\infty dt'} 
\def\intk{\int_{\bf k}} 
\def\intkd{\int \frac{d^dk}{(2\pi)^d}}
\def\intq{\int_{\bf q}} 
\def\intr{\int d^dr}  
\def\dintr{\displaystyle \int d^dr} 
\def\intrp{\int d^dr'}
\def\dinttau{\displaystyle \int_0^\beta d\tau}
\def\dinttaup{\displaystyle \int_0^\beta d\tau'}
\def\inttau{\int_0^\beta d\tau}
\def\inttaup{\int_0^\beta d\tau'}
\def\intx{\int d^{d+1}x} 
\def\inttaur{\int_0^\beta d\tau \int d^dr}
\def\intinf{\int_{-\infty}^\infty}
\def\dinttaur{\displaystyle \int_0^\beta d\tau \int d^dr}
\def\dintinf{\displaystyle \int_{-\infty}^\infty}
\def\intw{\int_{-\infty}^\infty \frac{d\w}{2\pi}}
\def\sumr{\sum_{\bf r}} 

\def\calA{{\cal A}} 
\def\calC{{\cal C}} 
\def\dt{\partial_t}
\def\calD{{\cal D}}
\def\calF{{\cal F}} 
\def\calG{{\cal G}}
\def\calH{{\cal H}}
\def\calI{{\cal I}}
\def\calJ{{\cal J}}
\def\calK{{\cal K}}
\def\calL{{\cal L}} 
\def\calN{{\cal N}}
\def\calO{{\cal O}}
\def\calP{{\cal P}}  
\def\calR{{\cal R}} 
\def\calS{{\cal S}}
\def\calT{{\cal T}}
\def\calU{{\cal U}}
\def\calY{{\cal Y}} 
\def\calZ{{\cal Z}} 

\def\calFbf{{\bf F}}

\def\tT{{\tilde T}}
\def\talpha{{\tilde\alpha}}
\def\tdelta{{\tilde\delta}}
\def\teta{{\tilde\eta}} 
\def\tlamb{{\tilde\lambda}}
\def\tmu{{\tilde\mu}}
\def\tphibf{{\tilde\phibf}}
\def\trho{{\tilde\rho}}
\def\tvarphibf{{\tilde\varphibf}} 
\def\tw{{\tilde\omega}}
\def\twn{{\tilde\omega_n}}

\def\asinh{{\rm asinh}} 
\graphicspath{{./figures_submit/}}

\def\Tkt{T_{\rm BKT}}
\newcommand{\calFQO}[1]{\calF_{\rm Qu-XY}^{(#1)}}

\section{Introduction}

Understanding the various phases of matter and the transitions between them is one of the main goals of condensed-matter physics. Of particular interest are (continuous) quantum phase transitions between different ground states, driven by quantum fluctuations related to Heisenberg's uncertainty principle (for reviews, see Refs.~\cite{Sondhi97,Coleman05,Sachdev11,Sachdev_book}). Although these transitions occur at zero temperature, the quantum critical point (QCP) controls the behavior of the system in a wide temperature range and often leads to intriguing physical properties with no equivalent in well-known phases of matter.  

A key concept for understanding quantum phase transitions is quantum criticality. Near a QCP, the system shows a universal scaling behavior which manifests itself not only in universal critical exponents but also in universal scaling functions. While understanding quantum criticality in strongly correlated systems is often a challenge, both experimentally and theoretically, cold atoms offer clean systems for a quantitative and precise study of quantum phase transitions. Quantum criticality in cold atoms has attracted increasing theoretical interest in the last years~\cite{Zhou10,Hazzard11,Fang11,Rancon12a}.

In ultracold gases, strong correlations can be achieved by tuning the atom-atom interactions by means of a Feshbach resonance, or by loading the atoms into an optical lattice~\cite{Bloch08}. In the latter case, by varying the strength of the lattice potential, it is possible to induce a quantum phase transition between superfluid and Mott insulating ground states in a Bose gas~\cite{Jaksch98,Greiner02}. The main features of this transition can be understood in the framework of the Bose-Hubbard model, which describes bosons moving in a lattice with an on-site repulsive interaction~\cite{Fisher89}. The density-driven Mott transition belongs to the same universality class as the transition between the vacuum and the superfluid phase in a dilute Bose gas~\cite{Fisher89,Sachdev_book}. When the transition occurs at fixed density, and is driven by a change in the interaction strength, it belongs to the $(d+1)$-dimensional XY universality class ($d$ is the space dimension). 

We have recently studied the equation of state near the three-dimensional density-driven Mott transition~\cite{Rancon12a,Rancon12d}. In this Letter, we focus on the interaction-driven Mott transition in a two-dimensional Bose gas. First, we review the critical behavior at the Mott transition in the Bose-Hubbard model, distinguishing between the generic QCP (density-driven transition) and the quantum multicritical point (QMCP) where the transition occurs at fixed density. In both cases, we write the equation of state using a universal scaling function and a small number of nonuniversal parameters whose value depends on the microscopic parameters of the Bose-Hubbard model. We then discuss in more detail the equation of state in the vicinity a QMCP using a nonperturbative renormalization-group (NPRG) approach. In particular, we show that the scaling function is similar to the one that was recently obtained within the two-dimensional quantum O(2) model~\cite{Rancon13a}. We conclude with a discussion of the 
experimental observation of quantum XY criticality 
in a two-dimensional Bose gas in an optical lattice.

\section{Critical behavior at the Mott transition} 
\label{sec_critical} 

The $d$-dimensional Bose-Hubbard model~\cite{Fisher89} is defined by the (Euclidean) action 
\begin{align}
S = \inttau \biggl\lbrace & \sum_\r \Bigl[ \psi_\r^* (\dtau-\mu)\psi_\r + \frac{U}{2} (\psi_\r^*\psi_\r)^2 \Bigr] \nonumber \\ & - t \sum_{\mean{\r,\r'}} \left(\psi_\r^* \psi_{\r'}+\mbox{c.c.}\right) \biggr\rbrace ,
\label{action}
\end{align}
where $\psi_\r(\tau)$ is a complex field and $\tau\in [0,\beta]$ an imaginary time with $\beta=1/T$ the inverse temperature. $\lbrace\r\rbrace$ denotes the $N$ sites of the lattice which is assumed hypercubic, $U$ the on-site repulsion, and $t$ the hopping amplitude between nearest-neighbor sites $\mean{\r,\r'}$. We set $\hbar=k_B=1$ and denote by $l$ the lattice spacing. 

\begin{figure}
\centerline{\includegraphics[width=7cm]{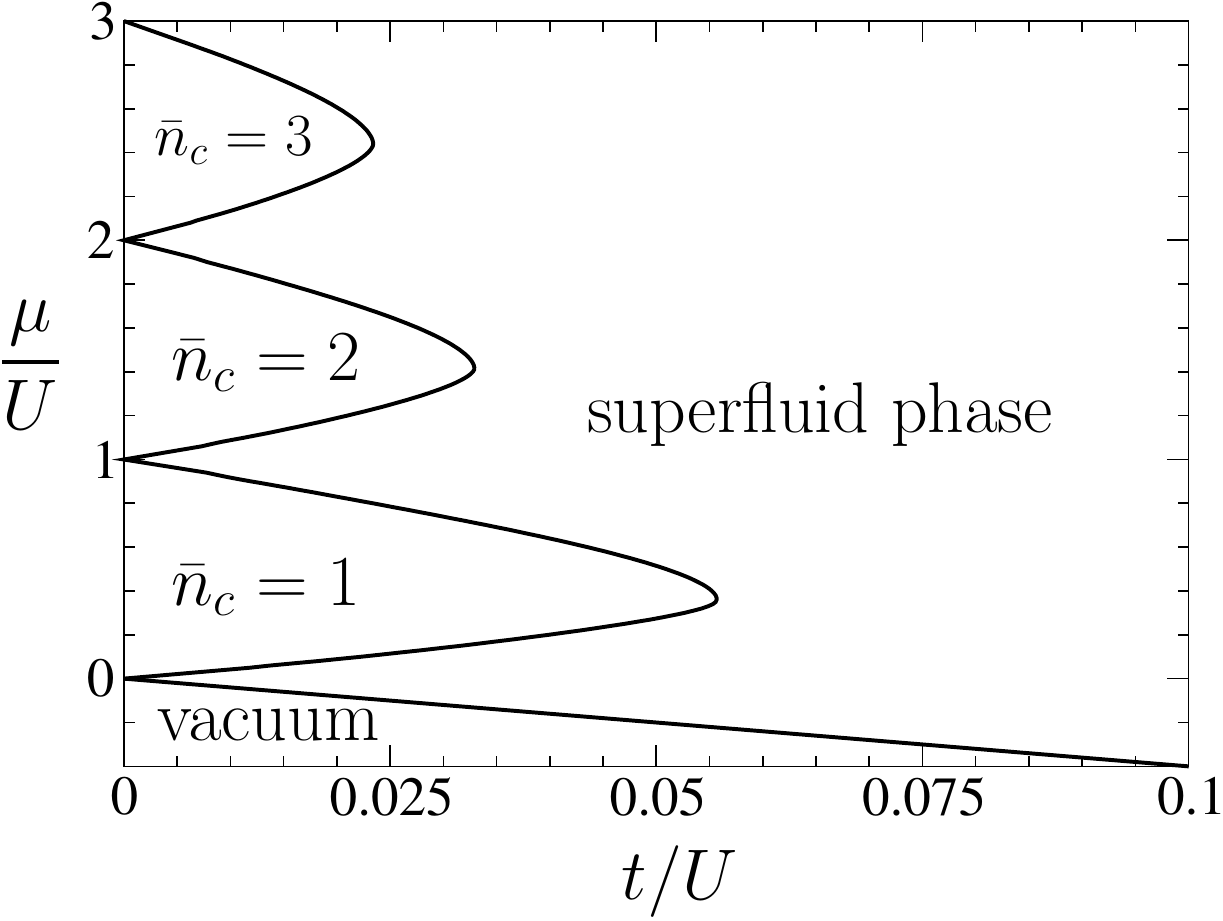}}
\caption{Zero-temperature phase diagram of the two-dimensional Bose-Hubbard model on a square lattice showing the first three Mott lobes, the vacuum and the superfluid phase.}
\label{fig_phase_dia}
\end{figure}

The zero-temperature phase diagram of the Bose-Hubbard model, obtained from the NPRG, is shown in Fig.~\ref{fig_phase_dia}~\cite{not7} (for $d=2$). For large $t/U$, the ground state is always superfluid for $\mu>-2dt$ (when $\mu\leq -2dt$ there are no particles in the system). For small values of $t/U$, one observes a series of lobes corresponding to incompressible Mott insulating phases with a commensurate density (integer mean number of bosons per site). The quantum phase transition between the superfluid phase and the Mott insulator belongs to two universality classes~\cite{Fisher89}. At the tips of the Mott lobes, where the transition takes place at constant density and is interaction driven, the universality class is that of the $(d+1)$-dimensional XY model with a dynamical critical exponent $z=1$. Anywhere else, the transition is driven by a density change and is in the dilute Bose gas universality class with a dynamical critical exponent $z=2$. 

To understand in more detail the thermodynamic properties of the system in the vicinity of the Mott transition, we introduce the effective action 
\beq
\Gamma[\phi^*,\phi] = - \ln Z[J^*,J] + \inttau \sum_\r [J_\r^* \phi_\r + \cc ] 
\label{Gammadef}
\eeq
defined as the Legendre transform of the thermodynamic potential $-\ln Z[J^*,J]$ ($Z[J^*,J]$ denotes the partition function), where $J_\r$ is a complex external source that couples linearly to the bosonic field $\psi_\r$ and $\phi_\r(\tau)=\delta \ln Z[J^*,J]/\delta J_\r^*(\tau)$ the superfluid order parameter. Thermodynamic properties of the system can be derived from the effective potential $V(n)=(\beta N)^{-1}\Gamma[\phi^*,\phi]$ with $n=|\phi|^2$ and $\phi$ a uniform and time-independent field. The minimum of $V(n)$ determines the condensate density $n_0$ and the pressure $P(\mu,T)=-V(n_0)$ in the equilibrium state. 

The critical behavior can be obtained from the low-energy expansion 
\begin{multline}
\Gamma[\phi^*,\phi] = l^{-d} \inttau \intr \Bigl[ \phi^* ( Z_C \dtau - V_A \partial_\tau^2 - Z_A t \nablabf^2 ) \phi \\  + V(n_0) + \delta(n-n_0)+  \frac{\lamb}{2} (n-n_0)^2  + \cdots \Bigr] ,
\label{Gamma}
\end{multline} 
where the ellipses denote higher-order (in derivative or field) terms. Equation~(\ref{Gamma}) is obtained by expanding the effective potential $V(n)$ about $n_0$ and retaining only the lowest-order derivative terms. We have taken the continuum limit where $\r$ becomes a continuous variable. At zero temperature, $\delta$ is nonzero in the Mott insulator and vanishes in the superfluid phase (where $n_0>0$) so that the transition line is given by $\delta\equiv \delta(t,\mu,U)=0^+$. $\delta$ and $Z_C$ are not independent but satisfy the relation~\cite{Fisher89,Sachdev_book,Rancon11b}  
\beq
Z_C = - \frac{\partial \delta}{\partial \mu} \biggl|_{t,U} ,
\eeq
which follows from the invariance of the action~(\ref{action}) in the semilocal U(1) transformation $\psi_\r(\tau)\to \psi_\r(\tau) e^{i\theta(\tau)}$ and $\mu\to \mu+i\dtau\theta(\tau)$ (with $\theta(\tau)$ a uniform time-dependent phase). 

Away from the tips of the Mott lobes, $Z_C$ is therefore nonzero and the dynamical critical exponent takes the value $z=2$. The second-order time derivative term in Eq.~(\ref{Gamma}) is then subleading and can be ignored; the transition is in the universality class of the dilute Bose gas and the upper critical dimension is $d_c^+=2$~\cite{Sachdev_book,Fisher89}. The similarity with the vacuum-superfluid transition of the dilute Bose gas can be made more explicit by introducing the effective mass $m^*$ of the critical fluctuations $\w=\q^2/2m^*$ at the QCP, as well as an effective scattering length $a^*$ characterizing the low-energy behavior of the interaction $\lamb$ between elementary excitations. In the vicinity of the QCP and for $d\geq 2$, the pressure reads~\cite{Rancon12d}
\begin{align}
P(\mu,T) ={}& P_c + \bar n_c \delta\mu \nonumber \\ 
& + \left(\frac{m^*}{2\pi}\right)^{d/2} T^{d/2+1} \calF_{\rm DBG}^{(d)} \left(\pm \frac{\delta\mu}{T}, \tilde g(T) \right) ,
\label{Pdbg} 
\end{align}
where $P_c$ and $\bar n_c$ denotes the pressure and the mean density at the QCP, respectively. $\calF_{\rm DBG}^{(d)}$ is a universal scaling function characteristic of the $d$-dimensional dilute Bose gas universality class. $\delta\mu=\mu-\mu_c$ measures the distance to the QCP and the dimensionless interaction constant $\tilde g(T)$ is a function of $m^*{a^*}^2T$. The $+$ ($-$) sign in Eq.~(\ref{Pdbg}) corresponds to particle (hole) doping of the Mott insulator. The scaling form~(\ref{Pdbg}) near a generic QCP has recently been verified in the three-dimensional Bose-Hubbard model, and the nonuniversal parameters $m^*$ and $a^*$ have been computed as a function of $t/U$ and $\mu/U$~\cite{Rancon12a,Rancon12d}.

At the tip of a Mott lobe, where the tangent to the transition line is vertical (Fig.~\ref{fig_phase_dia}), $Z_C$ vanishes and the dynamical critical exponent takes the value $z=1$. The QMCP is then similar to the critical point of the $(d+1)$-dimensional XY model. The lower and upper critical dimensions are therefore $d_c^-=1$ and $d_c^+=3$, respectively. The zero-temperature transition from the Mott insulator to the superfluid phase is driven by the vanishing of the particle-hole excitation gap, while the density is conserved. The critical behavior as we move away from the QMCP can be understood from the singular part of the effective potential. When $Z_C$ vanishes~\cite{not1}, the zero-temperature phase transition is controlled by the fixed point of the $(d+1)$-dimensional XY model. There is one relevant variable (that we denote by $r$) with scaling dimension $1/\nu$ given by the correlation-length exponent $\nu\equiv \nu_{\rm XY}^{(d+1)}$ of the $(d+1)$ dimensional XY model. If we move away from the Mott 
lobe tip in an arbitrary 
direction, $Z_C$ will in general not vanish. Denoting by $y$ its scaling dimension, the singular part of the effective potential satisfies, when $d<d_c^+$, the hyperscaling relation~\cite{Fisher89}
\begin{align}
V_{\rm sing}(r,Z_C) &= s^{-d-z}  V_s(s^{1/\nu}r,s^y Z_C) \nonumber \\ 
& \sim |r|^{\nu(d+z)} \tilde V_{\rm sing}\left( \frac{Z_C}{|r|^{y\nu}} \right) .
\label{Vsing}  
\end{align} 
The last result in~(\ref{Vsing}) is obtained with $s\sim |r|^{-\nu}$. $V_{\rm sing}$ being finite in the limit $Z_C\to 0$, $\tilde V_{\rm sing}(x)$ must behave as a constant in the limit $x\to 0$. Moreover $r$ and $Z_C$ are presumably analytic functions of $t-t_c$ and $\mu-\mu_c$, and must vanish linearly with $t-t_c$ as we approach a QMCP $(t_c,\mu_c)$ on a typical path (i.e. a path which is not vertical in the $(t/U,\mu/U)$ plane~\cite{not2}). Since $y=1$~\cite{Fisher89,Rancon11b} and $1-\nu_{\rm XY}^{(d+1)}>0$ for all dimensions $d+1\geq 3$, the argument of $\tilde V_{\rm sing}$ in Eq.~(\ref{Vsing}) vanishes as $t-t_c\to 0$. Given that $\tilde V_{\rm sing}(x)\to \const$ as $x\to 0$, we conclude that $Z_C$ drops out of the scaling relation~(\ref{Vsing}) and the multicritical point looks like an ordinary $(d+1)$-dimensional XY critical point. At finite temperature, the singular part of the effective potential satisfies 
\beq
V_{\rm sing}(r,T) \sim |r|^{\nu(d+z)} \tilde W_{\rm sing}\left( \frac{T}{|r|^{z\nu}} \right) ,
\eeq
using the fact that the scaling dimension of the temperature is given by the critical dynamical exponent $z$.

These observations imply that the universal (critical) behavior in the vicinity of a QMCP can be obtained from the quantum O(2) model 
\begin{align}
S[\varphibf] ={}& \inttau \intr \biggl\lbrace \half (\nablabf\varphibf)^2 + \frac{1}{2c_0^2} (\dtau \varphibf)^2 \nonumber \\ & 
+ \frac{r_0}{2} \varphibf^2 + \frac{u_0}{4!} {(\varphibf^2)}^2 \biggr\rbrace ,  
\label{action1} 
\end{align}
where $\varphibf$ is a 2-component real field satisfying periodic boundary conditions $\varphibf(\r,\tau+\beta)=\varphibf(\r,\tau)$. Note that this model has no first-order time derivative and exhibits Lorentz invariance at zero temperature. There is a QCP for a critical value $r_{0c}$ of $r_0$ (considering $u_0$ fixed) separating a disordered phase ($r_0>r_{c0}$) from an ordered phase ($r_0<r_{0c}$) where the O(2) symmetry is spontaneously broken. In two dimensions, the system is always disordered at finite temperatures but exhibits a Berezinskii-Kosterlitz-Thouless (BKT) phase transition for $r<r_{0c}$. In the universal regime near the QCP the pressure reads
\beq
P(T) = P(0) + 2 \frac{T^{d+1}}{c^d} \calFQO{d}\left( \frac{\Delta}{T} \right) 
\label{Pscaling}
\eeq
for $d_c^-\leq d\leq d_c^+$, where $c$ is the velocity of the critical fluctuations at the QCP and $|\Delta|$ a characteristic zero-temperature energy scale~\cite{Rancon13a}. In the disordered phase ($r_0>r_{0c}$), $\Delta$ is equal to the excitation gap of the $\varphibf$ field. When $r_0<r_{0c}$, it is convenient to take $\Delta$ negative such that $|\Delta|=-\Delta$ is the excitation gap in the disordered phase at the point located symmetrically with respect to the QCP. The universal scaling function $\calF\equiv\calFQO{2}$ of the two-dimensional quantum O(2) model has recently been computed using the NPRG~\cite{Rancon13a}. 

The pressure in the two-dimensional Bose-Hubbard model, in the vicinity of a QMCP, is given by Eq.~(\ref{Pscaling}) if we identify $\Delta$ with the one-particle excitation gap in the Mott phase. In the following, we discuss the equation of state for a constant chemical potential $\mu_c$ in the vicinity of a QMCP $(t_c,\mu_c)$. The gap $\Delta=\alpha U[(t_c-t)/U]^{z\nu}$ in the Mott insulator can be expressed as a function of the distance $t_c-t$ to the QCP, where $\alpha$ is a nonuniversal dimensionless number which depends on the Mott lobe considered. 

To compute the effective action $\Gamma[\phi^*,\phi]$ and the pressure $P(\mu,T)$ in the Bose-Hubbard model, we use the NPRG. One considers a scale-dependent effective action $\Gamma_k[\phi^*,\phi]$ which includes fluctuations with momentum $|\q|\gtrsim k$ and coincides with the effective action~(\ref{Gammadef}) when the momentum scale $k$ vanishes. In practice, this is achieved by adding to the action~(\ref{action}) a ``regulator'' term $\Delta S_k$ which suppresses fluctuations with momentum $|\q|\lesssim k$. For $k$ equal to a microscopic scale $\Lamb$ (of the order of the inverse lattice spacing), the action $S+\Delta S_\Lamb$ describes a system of decoupled sites (vanishing hopping amplitude) and is exactly solvable. To obtain the effective action $\Gamma\equiv\Gamma_{k=0}$ of the Bose-Hubbard model from $\Gamma_\Lamb$, we use a RG equation $\dk\Gamma_k$. We refer to Refs.~\cite{Rancon11a,Rancon11b} for a detailed discussion of this method and the approximations used to solve the RG equation.   

It is useful, in particular regarding experiments in cold atoms, to determine the domain of validity of the scaling form~(\ref{Pscaling}) in the two-dimensional Bose-Hubbard model. This can be done by looking at the RG flow of the coupling constant $\lamb_k$~\cite{not5}. At the QMCP, we can clearly distinguish two regimes: i) a (high-energy) nonuniversal regime $k\gtrsim k_G$ where lattice effects are important and the dimensionless coupling constant $\tlamb_k$ varies strongly with $k$, ii) a universal (critical) regime $k\ll k_G$ where $\tlamb_k$ is close to its fixed-point value $\tlamb^*$. As shown by the numerical solution of the flow equations, the crossover Ginzburg scale $k_G$ is of order of the inverse lattice  spacing $l^{-1}$ (the Ginzburg length $k_G^{-1}$ is typically equal to a few lattice spacings $l$)~\cite{not6}. Away from the QMCP, the energy scale $|\Delta|$ and the temperature define two new momentum scales, $k_\Delta = |\Delta|/
c$ and $k_T = T/c$, where $c$ is the velocity 
of the critical 
fluctuations. $k_\Delta^{-1}$ is the correlation length in the zero-temperature Mott insulator, and corresponds to the Josephson length in the superfluid phase. Universality requires $k_\Delta,k_T\ll k_G$, i.e. $|\Delta|,T\ll ck_G$. If we approximate $c$ by its value in the strong-coupling random-phase approximation (RPA), we finally obtain 
\beq
|\Delta|, T \ll k_G l \sqrt{t_cU}(\bar n_c^2 + \bar n_c)^{1/4} ,
\label{crit} 
\eeq
where $\bar n_c$ is the mean boson density at the QMCP (and in the nearby Mott insulator). The crossover energy scale below which universality is expected is therefore determined by $\sqrt{t_cU}$. This should be compared with the crossover scale $\sim t_c$ which controls the universal behavior in the vicinity of a generic QCP~\cite{Rancon12d}.

\section{Equation of state near a multicritical point}
\label{sec_eos} 

\begin{figure}
\centerline{\includegraphics[width=6.5cm]{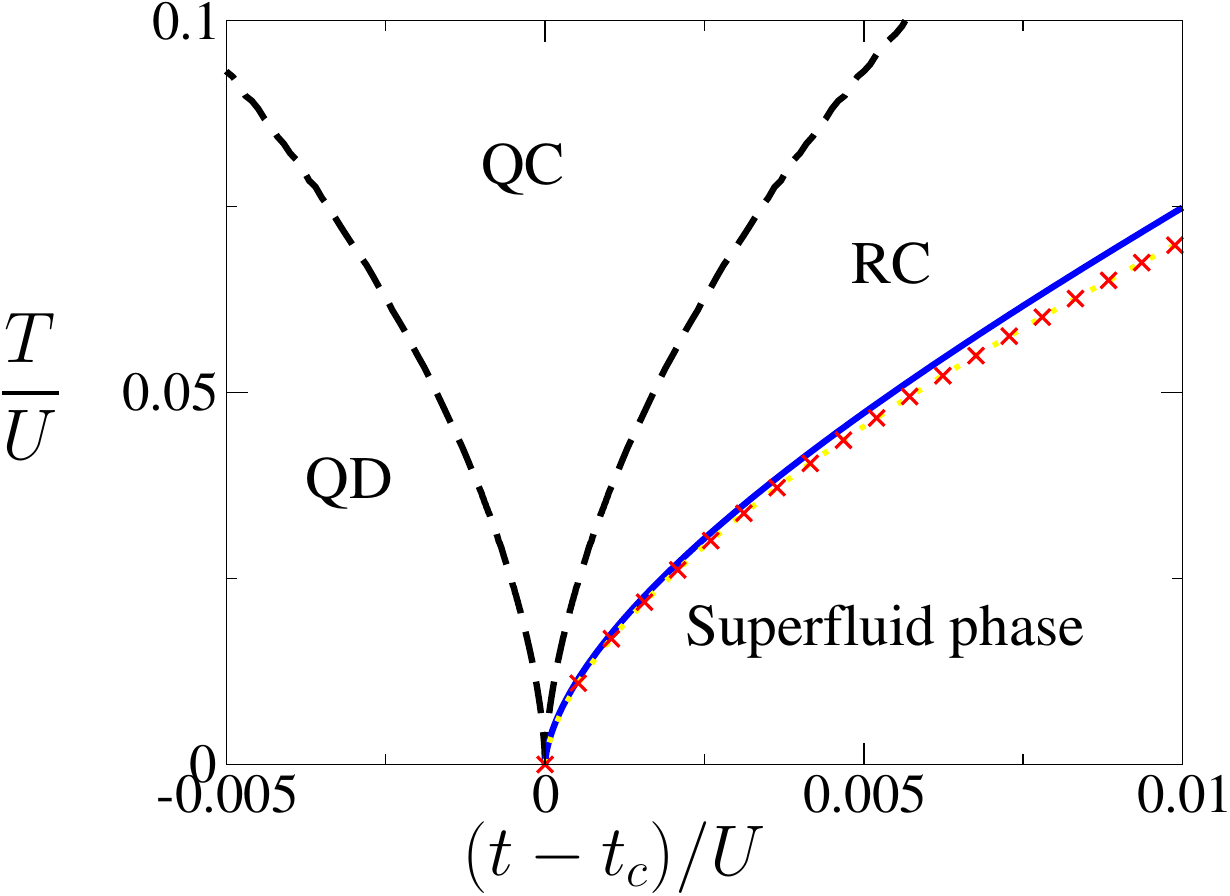}}
\caption{(Color online) Phase diagram near the multicritical point $(t_c,\mu_c)$ of the first Mott lobe for a constant chemical potential $\mu=\mu_c$ (QD: quantum disordered, QC: quantum critical, RC: renormalized classical). The (blue) solid line shows the BKT transition temperature obtained from $\Tkt=1.59\,\rho_s(T=0)$; the (red) crosses correspond to $T/U=1.28[(t-t_c)/U]^\nu$ (with $\nu=0.63$). The dashed crossover lines are obtained from the criterion $|\Delta|= T$.}
\label{fig_phase_dia_1} 
\end{figure}

The phase diagram near the QMCP $(t_c,\mu_c)$ of the first Mott lobe is shown in Fig.~\ref{fig_phase_dia_1}. At finite temperatures, we can distinguish three characteristic regimes by comparing $|\Delta|$ and $T$~\cite{Sachdev_book}: i) a quantum disordered regime ($\Delta\gg T$), a quantum critical regime ($|\Delta|\ll T$), and a renormalized classical regime ($-\Delta\gg T$). In the renormalized classical regime, there is a BKT phase transition between a high-temperature normal phase and a low-temperature superfluid phase with algebraic order. To estimate $\Tkt$, we use $\Tkt=\calC \rho_s$ where $\rho_s=\rho_s(T=0)$ is the zero-temperature superfluid stiffness (obtained in the NPRG approach) and $\calC$ a universal number close to $\pi/2$~\cite{not3} ($\Tkt$ in Fig.~\ref{fig_phase_dia_1} was obtained with $\calC=1.59$~\cite{Rancon13a}). Near  the QMCP the transition temperature is well approximated by $\Tkt/U\simeq 1.28[(t-t_c)/U]^{\nu}$, 
in very good agreement with the quantum Monte Carlo (QMC) result $\Tkt/U\simeq 1.29[(t-t_c)/U]^{\nu}$~\cite{Capogrosso08,not4}. The critical regime [Eq.~(\ref{crit})] is roughly defined by $T/U\lesssim 0.07$ and $|t-t_c|/U\lesssim 0.004$ if we take $k_Gl\sim 0.25$ (see below). 

\begin{figure}
\centerline{\includegraphics[width=7cm]{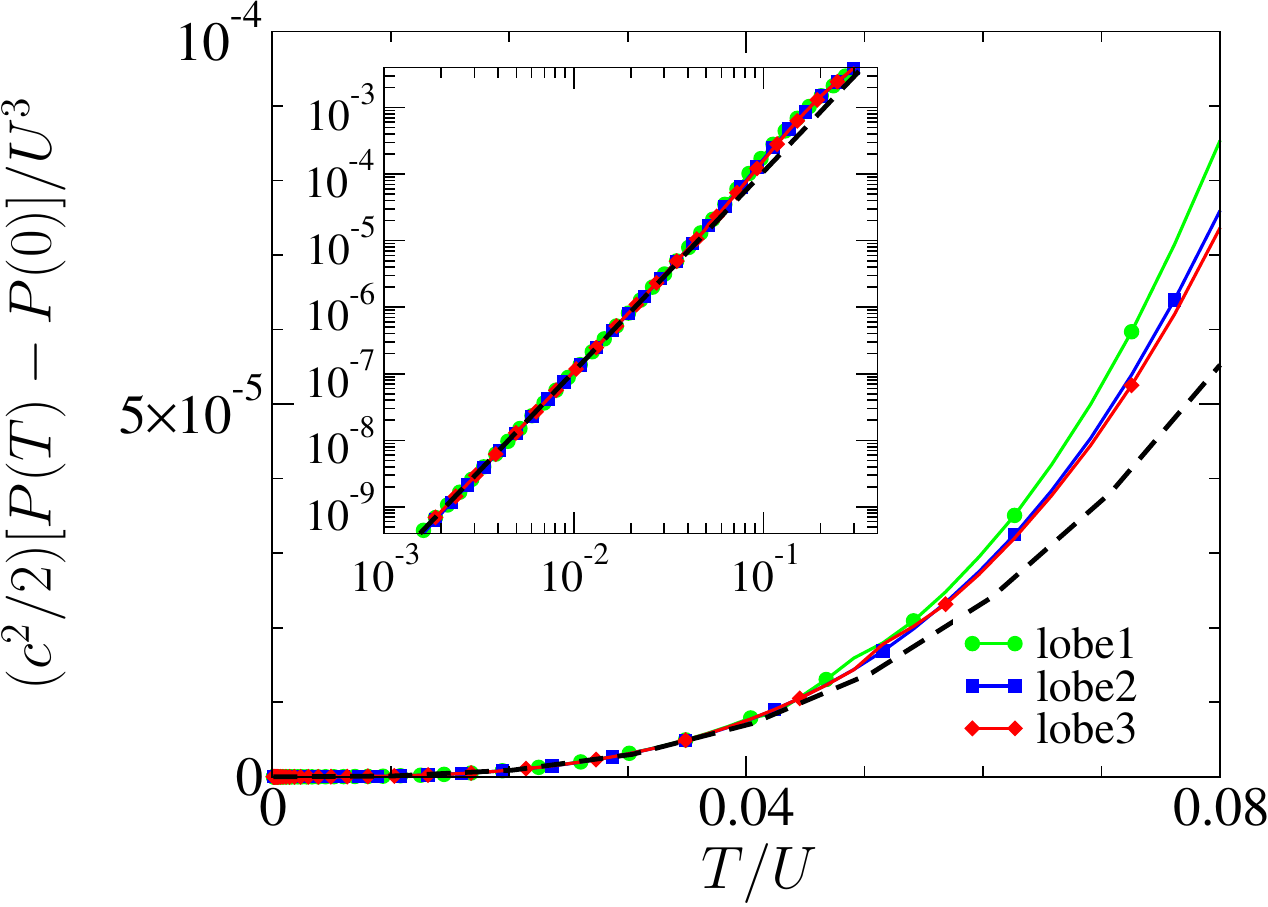}}
\caption{(Color online) Pressure {\it vs} temperature at the multicritical points of the first three Mott lobes. The dashed curve corresponds to $y=0.107\,x^3$. Inset: log-log plot showing the $T^3$ dependence of the pressure at low temperatures. }
\label{fig_pressure_qcp} 
\vspace{0.35cm}
\centerline{\includegraphics[width=7cm]{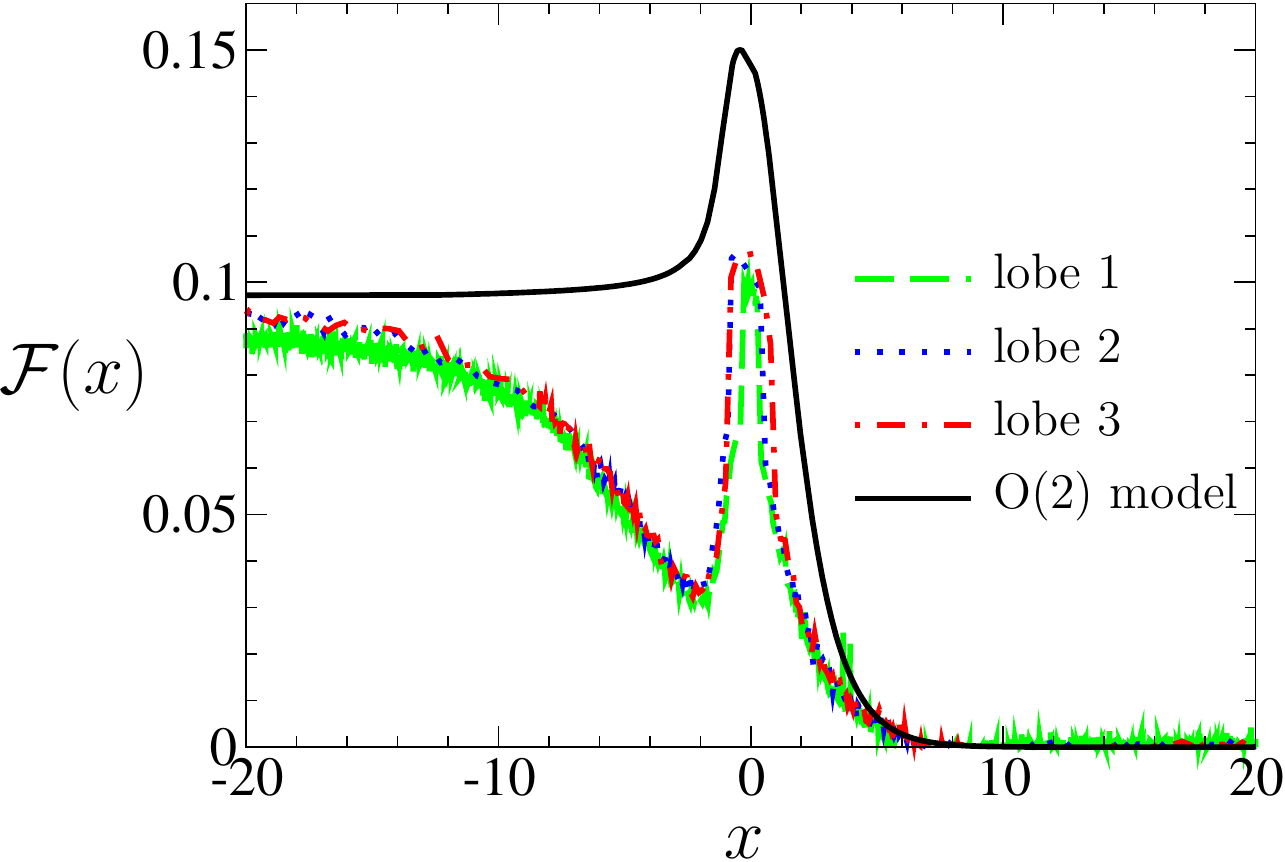}}
\caption{(Color online) Universal scaling function $\calF$ obtained from the pressure near the multicritical points of the first three Mott lobes. The (black) solid curve shows the result obtained from the two-dimensional quantum O(2) model~\cite{Rancon13a}.}  
\label{fig_F}  
\end{figure}

\begin{table}
\caption{Velocity $c$ and parameter $\alpha$ for the first three Mott lobes. The QMC data is taken from Ref.~\cite{Capogrosso08} and the velocity in the strong-coupling RPA is given by $c=\sqrt{t_cU}(\bar n_c^2+\bar n_c)^{1/4}$.}
\renewcommand{\arraystretch}{1.5}
\begin{center}
\begin{tabular}{lccc}
\hline \hline
Mott lobe & $\bar n_c=1$ & $\bar n_c=2$ & $\bar n_c=3$ \\
\hline
$c/lt_c$ NPRG & 4.88 & 8.53 & 12.14  \\
$c/lt_c$ QMC & $4.8\pm 0.2$ &  &  \\
$c/lt_c$ RPA & 5.74 & 9.85 & 13.89 \\ 
$\alpha$ & 2.238 & 3.374 & 4.222 \\
\hline \hline
\end{tabular}
\end{center}
\label{table}
\end{table}

Figure~\ref{fig_pressure_qcp} shows the temperature dependence of the pressure at the QMCP's of the first three Mott lobes. The corresponding values of the velocity $c$ and the parameter $\alpha$ are given in table~\ref{table}. We see that $(c^2/2)[P(T)-P(0)]$ varies as $T^3$ at low temperatures with a prefactor $\calF(0)\simeq 0.107$ which is independent of the multicritical point considered in agreement with the expected universality. The crossover temperature below which universality holds is of the order of $0.07U$ and agrees with the estimate~(\ref{crit}) if $k_Gl\sim 0.25$. 

Figure~\ref{fig_F} shows the full scaling function $\calF(x)$ as obtained from the first three Mott lobes. Again, the collapse of the data on a single curve is a convincing proof of universality. We do not obtain a perfect agreement with the scaling function derived from the quantum O(2) model~\cite{Rancon13a}, most likely for technical reasons related to the NPRG approach to the Bose-Hubbard model. The initial condition of the flow (namely the (local) limit of decoupled sites) makes it difficult to implement a RG approach which explicitly satisfies the emerging Lorentz invariance at the multicritical points. Even if the latter is very well satisfied by the RG flow~\cite{Rancon11b} a precise determination of $\calF$, yielding a better agreement with the result of the quantum O(2) model (in particular for the value of $\calF(0)$), appears difficult. Nevertheless our calculations reproduce the nonmonotonous behavior of the scaling function in the quantum critical regime with a peak in $\calF(x)$ located near 
$x=0$; they confirm that the QMCP's look like ordinary quantum XY 
critical points.

\section{Experimental observation} 
\label{sec_exp} 

Although cold atomic gases are inhomogeneous and of finite size due to the harmonic confining potential, from a local density approximation it is possible to deduce the pressure $P(\mu,T)$ of the infinite homogeneous gas (with uniform density) from the {\it in situ} density distribution $n(\r)$~\cite{Cheng07,Ho09}. This technique has recently been used to obtain the equation of state of a two-dimensional Bose gas in an optical lattice near the vacuum-superfluid transition~\cite{Zhang12}. The location of the QCP, as well as the critical exponents $z=2$ and $\nu=1/2$, were determined by writing the equation of state measured at various temperatures in a scaling form. The experimental results are in good agreement with a theoretical analysis of the two-dimensional Bose-Hubbard model~\cite{Rancon12b}. 

A similar experimental approach can be used to study the equation of state of a two-dimensional Bose gas in an optical lattice  near the Mott transition. The most direct evidence for quantum XY criticality in the vicinity of a multicritical point would come from a $T^3$ dependence of the pressure at low temperatures (at the generic Mott transition, the pressure varies quadratically with $T$). Because of three-body collisions, it is difficult to have a stable gas with more than two atoms per site in the Mott insulator~\cite{not8}. It is thus possible to measure the temperature dependence of the pressure at two different QMCP's (corresponding to the first two lobes). If we assume that the theoretical values of the velocity $c$ are accurate (as suggested by the agreement between NPRG and QMC for the first Mott lobe), we can then deduce the value of the universal number $\calF(0)$ and verify the prediction of the NPRG approach $\calF(0)\simeq 0.147$ obtained from the quantum O(2) model~\cite{Rancon13a}. 

A more ambitious goal consists in measuring the equation of state as a function of $t$ for a value of the chemical potential corresponding to a QMCP. A collapse of the data in agreement with the scaling form~(\ref{Pscaling}) would not only locate the position $(t_c,\mu_c)$ of the QMCP but would also yield an estimate of the critical exponents $\nu$ and $z$ as well as the full scaling function $\calF(x)$. 

One of the experimental difficulties in observing quantum XY criticality is that it requires to measure the pressure (i.e. the {\it in situ} density distribution $n(\r)$) at sufficiently low temperatures $T\lesssim 0.07U$ ($\sim t_c$ for the first Mott lobes).

\section{Conclusion} 

We have shown that the pressure $P(\mu,T)$ of a two-dimensional Bose gas near the interaction-driven Mott transition takes a universal form, with a universal scaling function $\calF\equiv\calFQO{2}$ characteristic of the quantum XY model universality class in two space dimensions [Eq.~(\ref{Pscaling})]. The nonuniversal parameters entering the equation of state, namely the velocity of the critical fluctuations and the parameter $\alpha$ relating the Mott gap $\Delta$ to the distance $t_c-t$ to the QMCP, have been computed for the first three Mott lobes (table~\ref{table}). Recent experiments have shown that it is now possible to measure the equation of state of a Bose gas in an optical lattice~\cite{Zhang12}. This opens up the possibility of a detailed study of the quantum XY model universality class in two dimensions, and in particular the determination of the universal scaling function $\calFQO{2}$. 

Recent theoretical works have focused on the amplitude (``Higgs'') mode which is expected in the vicinity of the QMCP's as a result of Lorentz invariance (or XY symmetry)\cite{Podolsky12,Podolsky11,Pollet12,Gazit13,Chen13}. The confinement of the gas by the harmonic trap suppresses the Higgs resonance which is replaced by a broad maximum in the spectral function. It is nevertheless possible to (indirectly) determine the energy of the Higgs mode from the onset of a strong response~\cite{Pollet12} as done in a recent experiment~\cite{Endres12}. Measuring the equation of state would provide us with complementary information as well as a more direct proof of the XY symmetry (or Lorentz invariance) at a QMCP between the superfluid phase and the Mott insulator.

\acknowledgments
A.R. acknowledges useful discussions with C. Chin and E. Hazlett.


\begin{thebibliography}{10}
\bibitem{not7}
The phase diagram was obtained using a simple form of the RG equations and is
  less precise than the one deduced from (more complicated) functional
  equations~\cite{Rancon11a,Rancon11b}. For the same reason, the value of the
  critical exponent $\nu\simeq 0.63$ at a QMCP is not as accurate as the one
  reported in Ref.~\cite{Rancon11b}.
\bibitem{not1}
In the close vicinity of a QMCP $(t_c,\mu_c)$, we expect $Z_C$ to vanish on a
  line roughly given by $\mu=\mu_c$.
\bibitem{not2}
For a vertical path ($t=t_c$), $r$ does not change sign and must therefore
  vanish as $(\mu-\mu_c)^2$~\cite{Fisher89}.
\bibitem{not5}
  In the low-energy limit, the scale-dependent effective action
  $\Gamma_k[\phi^*,\phi]$ can be expanded as in Eq.~(\ref{Gamma}) but with
  $k$-dependent parameters ($n_{0,k}$, $\delta_k$, $\lamb_k$,
  etc.)~\cite{Rancon11b}.
\bibitem{not3}
The BKT transition temperature is related to the jump of the stiffness by
  $T_{\rm BKT}=(\pi/2)\rho_s(T_{\rm BKT}^-)$~\cite{Nelson77a}. The fact fact
  $\calC=T_{\rm BKT}/\rho_s\simeq \pi/2$ (with $\rho_s\equiv \rho_s(T=0)$) is
  due to $\rho_s(T_{\rm BKT}^-)$ differing only slightly from
  $\rho_s$~\cite{Rancon13a}.
\bibitem{not4}
This expression is deduced from $\Tkt/t=0.49(U/t_c-U/t)^\nu$ with
  $t_c/U=0.05974$ and $\nu=0.6715$~\cite{Capogrosso08}.
\end{thebibliography}

\end{document}